%% file: GLSVLSI-2023.tex
\documentclass[conference]{_sty/IEEEtran}
\IEEEoverridecommandlockouts

\usepackage{algorithmic}
\usepackage{amsmath,amssymb,amsfonts}
\usepackage{array}
\usepackage{booktabs}
\usepackage{cite}
\usepackage{caption}
\usepackage{color}
\usepackage{comment}
\usepackage[inline]{enumitem}
\usepackage{epsfig}
\usepackage{etoolbox}
\usepackage{fancybox}
\usepackage{float}
\usepackage{stfloats}
\usepackage{fixltx2e}
\usepackage{graphicx}
\usepackage[breaklinks,colorlinks]{hyperref}
\usepackage{mathrsfs}
\usepackage{multirow}
\usepackage{multicol}
\usepackage{placeins}
\usepackage{rotating}
\usepackage{setspace}
\usepackage[hang, tight]{subfigure}
\usepackage{textcomp}
\usepackage{threeparttable}
\usepackage{times}
\usepackage{url}
\usepackage{verbatim}
\usepackage{wrapfig}
\usepackage[usenames,dvipsnames]{xcolor}
\usepackage{ulem}
\usepackage{cite}
\usepackage{amsmath,amssymb,amsfonts}
\usepackage{algorithmic}
\usepackage{graphicx}
\usepackage{textcomp}
\usepackage{xcolor}
\def\BibTeX{{\rm B\kern-.05em{\sc i\kern-.025em b}\kern-.08em
    T\kern-.1667em\lower.7ex\hbox{E}\kern-.125emX}}

\usepackage[capitalize]{cleveref}
\crefname{section}{Sec.}{Secs.}
\Crefname{section}{Section}{Sections}
\Crefname{table}{Table}{Tables}
\crefname{table}{Tab.}{Tabs.}

\newcommand{\tabincell}[2]{\begin{tabular}{@{}#1@{}}#2\end{tabular}}
\newcommand{\roma}[1]{\uppercase\expandafter{\romannumeral #1\relax}}
\graphicspath{{_fig/}}

\begin{document}

\title{EdgeShield: A Universal and Efficient Edge Computing Framework for Robust AI \vspace{-2mm}}

\author{\IEEEauthorblockN{Duo Zhong}
\IEEEauthorblockA{\textit{Computer Science and Electrical Engineering Department} \\
\textit{University of Maryland, Baltimore}\\
Baltimore, USA \\
duoz1@umbc.edu}
\and
\IEEEauthorblockN{Bojing Li}
\IEEEauthorblockA{\textit{Computer Science and Electrical Engineering Department} \\
\textit{University of Maryland, Baltimore}\\
Baltimore, USA \\
ji18978@umbc.edu}
\and
\IEEEauthorblockN{Xiang Chen}
\IEEEauthorblockA{\textit{Department of Electrical and Computer Engineering} \\
\textit{George Mason University}\\
Fairfax, USA \\
xchen26@gmu.edu}
\and
\IEEEauthorblockN{Chenchen Liu}
\IEEEauthorblockA{\textit{Computer Science and Electrical Engineering Department} \\
\textit{University of Maryland, Baltimore}\\
Baltimore, USA \\
ccliu@umbc.edu}
}

\maketitle

\input{_txt/0-Abstract_v2}
\input{_txt/1-Introduction_v5}
\input{_txt/2-Background_v4}

\input{_txt/3-Method_v3}

\input{_txt/4-Experiment_v5}
\input{_txt/5-Conclusion_v1}

\bibliographystyle{_sty/ieee}
\bibliography{_bib/Duo_references}



\vspace{12pt}
\color{red}

\end{document}

%% file: _txt/0-Abstract_v2.tex
\begin{abstract}
The increasing prevalence of adversarial attacks on Artificial Intelligence (AI) systems has created a need for innovative security measures. 
However, the current methods of defending against these attacks often come with a high computing cost and require back-end processing, making real-time defense challenging.
Fortunately, there have been remarkable advancements in edge-computing, which make it easier to deploy neural networks on edge devices. Building upon these advancements, we propose a edge framework design to enable universal and efficient detection of adversarial attacks. 
This framework incorporates an attention-based adversarial detection methodology and a lightweight detection network formation, making it suitable for a wide range of neural networks and can be deployed on edge devices.
To assess the effectiveness of our proposed framework, we conducted evaluations on five neural networks. The results indicate an impressive 97.43\% F-score can be achieved, demonstrating the framework's proficiency in detecting adversarial attacks.
Moreover, our proposed framework also exhibits significantly reduced computing complexity and cost in comparison to previous detection methods. This aspect is particularly beneficial as it ensures that the defense mechanism can be efficiently implemented in real-time on-edge devices.
\end{abstract}

%% file: _txt/1-Introduction_v5.tex
\section{Introduction}
Edge devices and sensors usually run in a complex and noisy environment, where data are exposed to contamination with a high possibility. Combating these contamination usually caused by adversarial attacks requires deep neural network (DNN) with specific anti-interference optimization.
However, the extremely limited computing resources of edge devices severely limit local deployment of high-workload robustness-strengthened neural networks.
As a result, given security concerns in practical applications, the centralised approach is still the most commonly used. 
In this mainstream mode, all the data collected by the edge sensor end will be transmitted to the central computing end (back end) regardless of perturbations and even attacks, and robustness-strengthened neural networks on the back end will shoulder the responsibility of fighting such adversarial attacks.
This creates redundant communication overhead and increases the workload on the computing center, which goes against the trend towards edge computing.

If we could isolate the task of combating adversarial attacks from robustness-strengthened neural networks and assign it to edge devices, this shift would considerably reduce communication overhead for tainted data would be filtered out at the source. Consequently, freed from the burden of security concerns, the back-end can reallocate more resource towards advanced computational tasks, optimizing the overall system efficiency and aligning with the goals of edge computing.


The adversarial attacks, especially adversarial patch attacks, a small yet carefully crafted patch is placed randomly on an image obtained by the sensor~\cite{eykholt2018robust, brown2017adversarial, karmon2018lavan} have emerged as an urgent threat~\cite{szegedy2013intriguing,goodfellow2014explaining,kurakin2018adversarial,moosavi2016deepfool,carlini2017adversarial,eykholt2018robust,brown2017adversarial,karmon2018lavan}. These patches have been proven to cause the back-end neural networks to generate abnormal high activation layer by layer~\cite{yu2021defending}, which weakens the authentic features of the original image and deceives the back-end system into misidentifying the image. The simplicity of training and the low-cost deployment of such attacks makes them attractive to malicious actors seeking vulnerabilities in visual intelligence systems. As a result, the trustworthiness and reliability of these systems are compromised.

Although many works have been proposed to tackle the security issue of intelligence systems caused by adversarial patch attacks, none of them are feasible to be deploy on edge devices. 
For example, Digital Watermarking (DW)~\cite{hayes2018visible} and Local Gradient Smoothing (LGS)~\cite{naseer2019local} were proposed as patch removal techniques that locate and erase (or mask) patches. 
LanCe~\cite{xu2020lance} defined Input Semantic Inconsistency for patch detection.
Yu~\cite{yu2021defending} introduced Feature Norm Clipping (FNC) layer between original model layers.
However, these approaches heavily rely on optimizing the neural networks on back end, which involves complex execution and large-scale deployment of these models. Such resource-intensive processes are not practical for edge devices, where computing power and memory capacity are limited.

As research findings have demonstrated, feature activation maps from attacked images typically exhibit abnormal higher values in some regions compared to those from original images. These abnormalities are considered low-level, or coarse features, which inspired us the thought to develop a specialized light-weight neural networks that can be easily deployed at resource constrained edge devices for their extraction. This aligns with our initial idea of transferring the task of combating adversarial attacks to edge devices. Moreover, this novel strategy can achieve real-time attacks detection without waiting for the back end processing, making a significant step forward in the responsiveness of AI applications.



\begin{figure*}[!t]
  \vspace{-3mm}
  \setlength{\abovecaptionskip}{1.4mm}
  \setlength{\belowcaptionskip}{-0.cm}
  \centering
  \includegraphics[width=7in, height=3.5in]{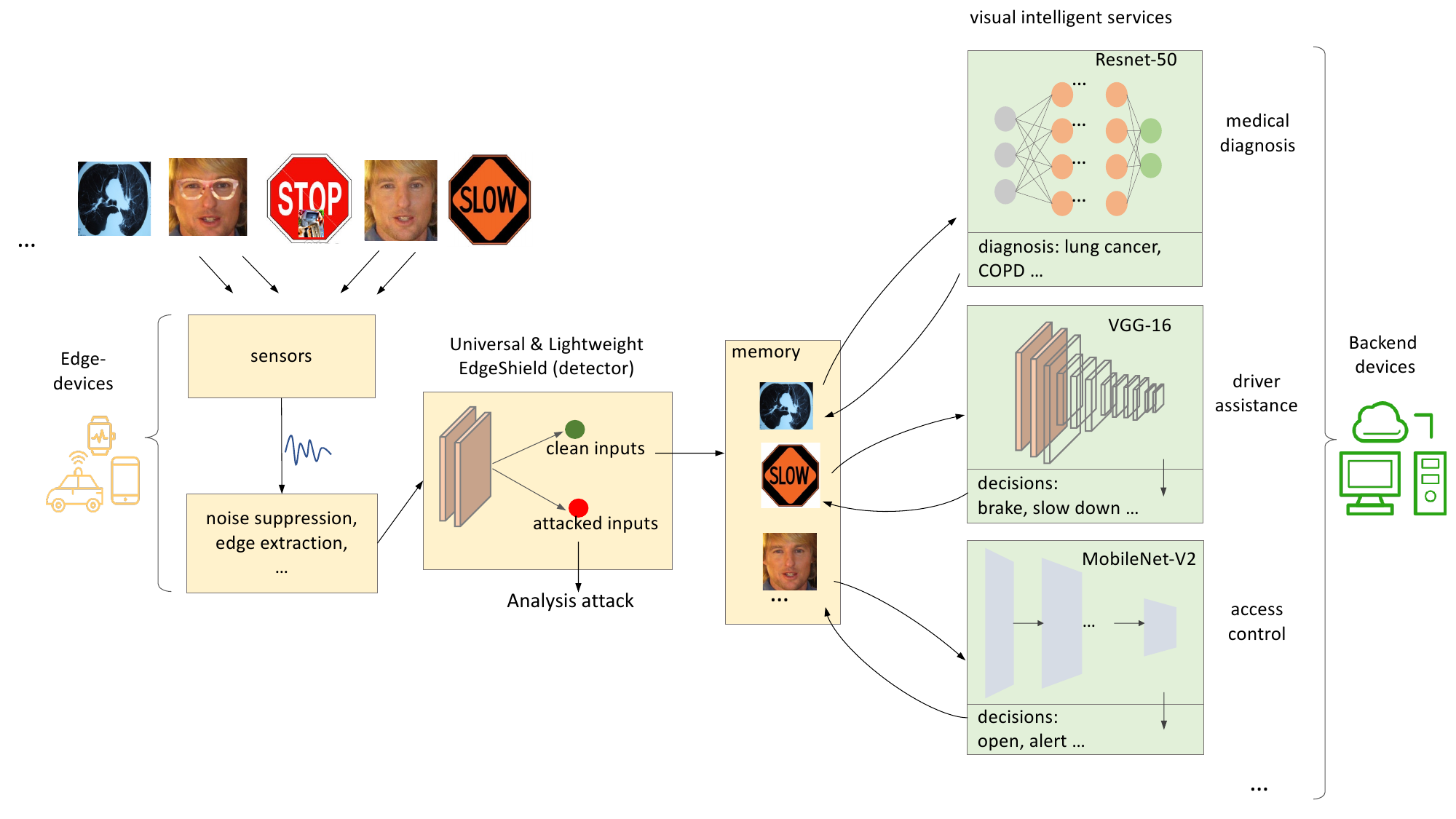}
  \caption{The framework of our proposed method.}
  \label{fig:Fig1}
\end{figure*}

In our work, we propose an edge computing framework that can effectively and efficiently detect abnormal data attacked by adversarial patches and purify sensory data.
As shown in Fig~\ref{fig:Fig1}, the main component of this framework is a lightweight yet versatile detection model.
This unified model is capable of detecting a wide range of adversarial patches that pose threats to most DNN models.
It is solely composed of a few shallow layers from the back end classical DNN models, which allows it to be deployed locally.
Additionally, we have developed an attention-based methodology to achieve accurate detection while introducing only simple computations.
By preventing contaminated data from being transmitted to the back-end cloud, our framework reduces communication bandwidth, enhances the accuracy, and reduces the workload of the back-end intelligence system. 



%% file: _txt/2-Background_v4.tex
\section{Background}
\subsection{AI Robustness}
The increasing advancements in DNNs have led to their widespread adoption in various AI systems, including but not limited to facial recognition, object detection, autonomous driving, and surveillance systems. The large-scale DNNs such as VGG-16~\cite{simonyan2014very}, ResNet-101~\cite{he2016deep}, MobileNet-V2~\cite{sandler2018mobilenetv2}, etc. 
are powerful with complex structures. 
However, these models are vulnerable to visual attacks, particularly adversarial patch attacks, where a small universal patch trained on a limited dataset is attached to the model's inputs~\cite{brown2017adversarial}. Patches trained against one model exhibit varying levels of transferability to other models. Therefore, this work focuses on developing a universal and light-weight model with low-cost detection methodology for detecting all adversarial patches trained against varying models. 

Although many approaches have been proposed to address this issue, they all suffer from certain limitations. DW~\cite{hayes2018visible} and LGS~\cite{naseer2019local} perform poorly in patch detection and can
compromise significant parts of the clean image. LanCe~\cite{xu2020lance} detects inconsistencies between the output of the model's last layer and the synthesized pattern of the predicted category. 
This approach requires computing thresholds for synthesized patterns of each category, which is resource- and time-consuming. 
Furthermore, the computation of the inconsistency requires knowledge of the predicted class of the output, limiting its use to post-prediction correction.
Yu et al.~\cite{yu2021defending} proposed adding FNC layers to clip the outputs of specific layers in the model. 
The FNC clipping strategy involves replacing the values of all points in the feature map where the L-2 norm is greater than the mean. 
However, this clipping strategy also affects the larger values in the feature map of clean images that contribute more to the classifier's output, resulting in noticeable damage to the clean images. 
Moreover, the introduction of FNC layers increase the algorithm complexity and the involved computations are resource expensive.
It is noted that these methods involve full neural network models, which also hinders their deployment on edge devices.


\subsection{Edge Computing}
Over the past decade, the swift advancements in microchip~\cite{webpage2023Snapdragon, webpage2023Movidius, webpage2023Apple} technology have markedly boosted edge computing in devices like mobile phones. These advancements have made it possible for low-level computational tasks, such as coarse feature extraction with lightweight neural networks, to run efficiently on edge devices, thereby streamlining processing and realizing real-time data handling. Furthermore, the introduction of frameworks, including CoreML~\cite{coremltools} and TensorFlow Lite~\cite{tensorflowlite}, has greatly simplified the deployment of DNNs on edge devices like iPhones and Raspberry Pi. This development has been instrumental in facilitating effective hardware-oriented performance optimization, making it more feasible to implement DNN capabilities in these compact, resource-limited devices. These two aspects progress set the stage for addressing more challenging tasks at the edge devices, such as detecting adversarial attacks. 

Isolating and shifting the task of detecting adversarial attacks to the edge is crucial. This shift reduces latency, allowing for quicker response to attacks. It also decreases the data bandwidth requirements between the edge and the back-end, enhancing overall system efficiency. Additionally, it supports scalable solutions, enabling edge devices to independently manage these security tasks, thus easing the load on central systems. This approach highlights the growing importance of edge computing in handling not only computational tasks but also critical security measures in an increasingly interconnected world.


\begin{figure}[t]
  \vspace{-3mm}
  \setlength{\abovecaptionskip}{1.4mm}
  \setlength{\belowcaptionskip}{-0.0cm}
  \centering
  \includegraphics[width=\linewidth]{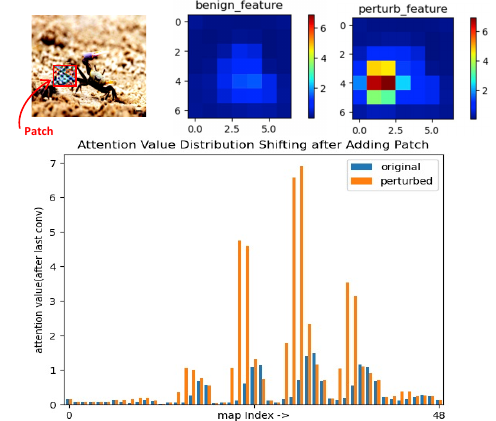}
  \caption{Attention values of clear and perturbed data.}
  \label{fig:Fig2}
  \vspace{-4mm}
\end{figure}

\begin{figure*}[t]
  \vspace{-3mm}
  \setlength{\abovecaptionskip}{1.3mm}
  \setlength{\belowcaptionskip}{-0.cm}
  \centering
  \includegraphics[width=\textwidth,height=1.4in]{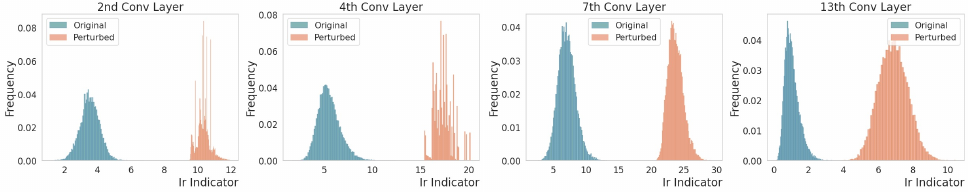}
  \caption{Distributions of $I_{r}$ indicators at different layers of VGG-16 on ImageNet.}
  \label{fig:Fig3}
  \vspace{-3mm}
\end{figure*}

%% file: _txt/3-Method_v3.tex
\section{Designed Methodology}

As shown in Fig.~\ref{fig:Fig1}, the proposed framework consists of two parts: detection and inference. 
The detection model is selected shallow layers of a neural network (e.g., VGG 16) and is deployed on an edge device that collects inputs through sensors. 
The detected adversarial images will be discarded and only clear images will be uploaded to the back-end computing centre for inference. 
The inference part is a full DNN deployed on the server such as VGG-16, ResNet-101, Inception-V3, etc. 

\subsection{Attention-based Method for Attacks Detection}
It is observed that adversarial patches mislead classifier result by causing some signiﬁcant abnormalities that overwhelm the original decision portion in the last convolution activation map~\cite{xu2020lance,yu2021defending}. Instead of introducing complex calculations like L-2 norm or inconsistency in previous works, we propose a low-cost attention-based methodology that uses the attention map generated from the activation map of a particular layer for detection. 
Each point in attention map is defined as Eq. 1.
\begin{equation}
A_{T}(F,h,w) = \frac{1}{C} \sum_{i}^{C}F_{h,w}(i)
\end{equation}
$F\in \mathbb{R}^{C\ast H\ast W}$ is the activation map of a certain layer in the model. $F_{h,w}(i)$ represents the activation value of the $i^{th}$ channel at coordinates \begin{math} (h,w)\end{math} in activation map. The attention map is calculated by taking the mean of the activation map along the channel dimension C. As such, an attention map $A_{T}(F)$ of size $H\ast W$ for the activation map $F$ is obtained.

Utilizing VGG-16 as an example, as shown in Fig.~\ref{fig:Fig2}, the attention maps generated by a specific activation layer show significant differences between clean and perturbed inputs. Specifically, the attention map for perturbed input exhibits a higher focus on the patch region. Inspired by this observation, we assume that the maximum value in the attention map contains the most salient features, and we define it as the indicator $I_{r}$:
\begin{equation}
I_{r}(A_{T}) = \max A_{T}(F,h,w)
\end{equation}

To leverage attention maps for adversarial detection, we first compute $I_{r}$ values for perturbed and their corresponding clean samples 
on a specific convolution layer of a DNN model. 
Then, we determine a threshold value $\Theta$ to distinguish the perturbed and clean samples with a confidence $p$, using only the $I_{r}$ values from clean samples.
\begin{equation}
p = \frac{\sum_{A_{T}\in A_{T}^{clean}}\mathbb I(I_{r}(A_{T})\leq \Theta)}{|A_{T}^{clean}|}
\end{equation}
$A_{T}^{clean}$ represents a set of attention maps generated from numerous clean samples. $|A_{T}^{clean}|$ indicates the number of elements in $A_{T}^{clean}$.
$\mathbb I(\cdot)$ is an indicator function that returns $1$ if the condition in parentheses is True, and $0$ otherwise. Given a confidence level $p$ (usually set to 0.95), we calculate a threshold value $\Theta$ such that $p$ fraction of the clean samples satisfy $I_{r}(A_{T})\leq \Theta$. It is worth noting that our method for computing $\Theta$ does not require any exposure to perturbed samples.

\begin{figure*}[!t]
  \setlength{\abovecaptionskip}{1.4mm}
  \setlength{\belowcaptionskip}{-0.cm}
  \centering
  \includegraphics[width=0.85\textwidth,height=2.5in]{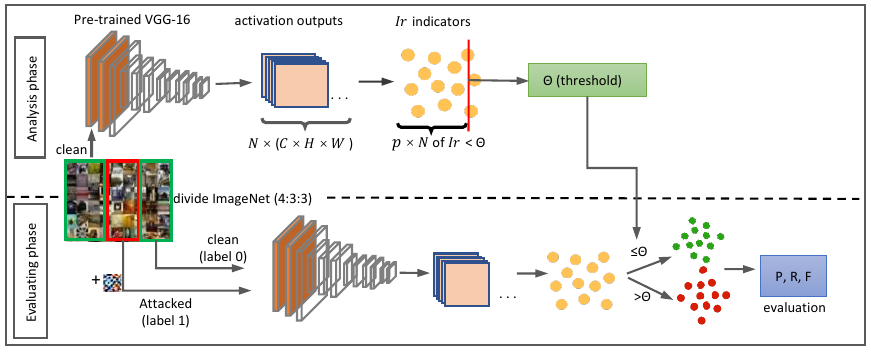}
  \caption{The implementation detail of the proposed detection.}
  \label{fig:Fig5}
  \vspace{-5mm}
\end{figure*}

Finally, the detection can be formulated as Eq. 4.
\begin{equation}
\mathcal{D}(x) = 
\begin{cases}
1\ \ (perturbed), \quad {if} \ \ {I_{r}(A_{T})>\Theta},\\
0\ \ (clean), \quad {otherwise.}
\end{cases}
\end{equation}
For an input sample $x$, we obtain its attention map $A_{T}$ and corresponding indicator $I_{r}$ on a specific convolution layer of a DNN model. 
Note that all models used in our study, including VGG-16, ResNet-101, Inception-V3, etc., do not require any additional training. 

\subsection{Detection Layer Selection}

Instead of involving the full DNN model to detection, we propose to select a few layer of a model to enable local deployment.
To determine the layers to be selected, $I_{r}$ indicators on various convolution layers of a DNN model is collected and their frequency histograms are shown in Fig.~\ref{fig:Fig3}. 
The results demonstrate that the distributions of clean and perturbed $I_{r}$ are more separate in shallower layers. This observation can be interpreted based on the principle of convolution operation and adversarial patches. 
Adversarial patches are designed to draw more attention to the areas where it placed of an image, which often results in increasing values in patches area of attention maps.
With this impact propagating to larger range pixels in deeper convolution layers to increase the overall impact, the increasing magnitude for individual points will inevitably weaken, i.e. the influence on upper bound of attention map will weaken.
Thus, searching the upper bound of the attention map at shallower layers make most sensitive way to keep track the significant differential, i.e. the indicator $I_{r}$ at shallower layers.

After analyzing the results mentioned above, we have confirmed that $I_{r}$ is a highly precise indicator for detecting attacks in shallow convolution layers. Consequently, selecting such layers as the detection layer can significantly decrease hardware costs during implementation. 
In addition, our approach enables real-time attack detection by eliminating the requirement of running the complete inference of DNNs for detection and only choosing 
a few shallow layers to form the detection model. 
For example, the detection model with the first two convolution layer of VGG-16 demonstrates optimal performance in the subsequent parts.

%% file: _txt/4-Experiment_v5.tex
\section{Experiment and Evaluation}

\subsection{Experiment Setup}
Fig.~\ref{fig:Fig5} indicates the implementation detail of the proposed detection. 
The framework includes two phases: the analysis, where the threshold $\Theta$ for a deployed DNN is calculated based on clean samples, and the evaluation, which assesses the detection performance.

\textbf{Experiment Dataset:} A random sampling of the ImageNet dataset~\cite{russakovsky2015imagenet} allocates 40\% for analysis and 60\% for testing purposes. Half of the test data are attacked to create a balanced set of positive (attacked) and negative (clean) samples. The analysis dataset is used to generate the threshold $\Theta$, derived from $I_{r}$ indicators, for detection deployment. Additionally, the effectiveness of this calculated threshold will be evaluated using the testing dataset.

\textbf{Detection Layer:} The shallow layers used to obtain $I_{r}$ indicators are from VGG-16. However, shallow layers from other classic models like ResNet-50, MobileNet-V3, etc. can also be employed. The selection of detector models and layers is discussed in the subsequent sections.

\textbf{Attacks Training:} We utilize the adversarial patch attacks technique in~\cite{brown2017adversarial} to train adversarial patches against each DNN separately, with each patch occupying only 6\% of the original image's pixels (54x54). The smaller the patch is, the more challenging its detection becomes. We take 6\% as is generally done in this study. A small sample of images (i.e., 4000) from ImageNet is used in the training stage. As shown in Table~\ref{tab:table1}, when inputs are compromised with these patches, the top-1 accuracy of the affected models on ImageNet can be even reduced to 0.83\% at most.

\input{_tab/tab1_v2}



\subsection{Effectiveness of Detection} 
Table~\ref{tab:table1} shows those adversarial patches cause varying accuracy degradation on their target DNNs. 
VGG-16 is the most vulnerable, while MoblieNet-V2 is relatively robust and difficult to attack.
In this work, we aim to build a unified model to effectively detect all those patches, no matter how their transferability are, to block them from fooling the subsequent DNNs on back end.

\input{_tab/tab2}

To achieve this goal, we take different shallow layers from VGG-16 to derive the indicator $I_{r}$ and calculate $\Theta$. 
The results presented in Table~\ref{tab:table2} indicate that taking the second convolutional layer of VGG-16 as the detection layer yields the most effective results, i.e., employing the  first two convolutional layers of VGG-16 as a unified detection model. This methodology consistently achieves a notable level of accuracy in detecting attacks, with the detection rate reliably reaching or exceeding 97.36\%.  
The results also show that the adversarial patches Advp-MobileNetV2, which is trained to attack the robust MobileNet-V2, tends to be more aggressive, making it challenging to detect. 
As such, the detection model can be formed with only the first two convolution layers of VGG-16, which is lightweight and can be deployed in the resource limited edge devices.
The results also indicate that in scenarios where computing resource is extremely limited, deploying only the first layer is still possible with an F-score of at least 87.91\%, even when subjected to the Advp-MoblieNetV2 attack.

\input{_tab/tab3}

We also conducted experiments to investigate the effectiveness of using shallow layers of different models as a unified model for detection. 
The results in Table~\ref{tab:table3} demonstrated that without considering patch Advp-MoblieNetV2, the first convolution layers of all models are suitable for forming the detection model, achieving an F-score of at least 92.17\%.
Considering Advp-MoblieNetV2, only the first convolution layers of ResNet-50 and MobileNet-V2 are appropriate for constructing the detection model, achieving an F-score of at least 80.91\%.

\textbf{Effectiveness Summary:} The results from Table~\ref{tab:table2} and Table~\ref{tab:table3} demonstrate that the unified detection model, taking the first two convolution layers from VGG-16 to obtain  indicator $I_{r}$ for detection, achieves the best performance against all adversarial patches trained under different threat models. Specifically, it attains an F-score of at least 97.42\%, significantly surpassing the detection accuracy of LanCe~\cite{xu2020lance}, with is 91\%.

\textbf{Interpretation of Results:} The optimal choice obtained from the experimental results may be attributed to the fact that VGG-16 is the most vulnerable model among all the threat models, as shown in Table~\ref{tab:table1}. This vulnerability allows patches trained by other threat models to possess some level of transferability to attack VGG-16. As a result, the shallow layers of VGG-16 demonstrate sensitivity to all patches and exhibit the capability to capture pertinent features that indicate the presence of adversarial patches.





\input{_tab/table8}
\subsection{Computing Efficiency}

The computing efficiency in terms of speed and energy consumption of the proposed detection model is evaluated across three platforms: GPU (NVIDIA GeForce RTX 3090), CPU$^{1}$ (12th Gen Intel(R) Core(TM) i9-12900KF), and CPU$^{2}$ (Apple M2).
We compared the efficiency of our detection model with that of the full models from which the detection model is derived.
The comparative analysis is conducted across three DNNs, whose shallow layers have demonstrate high detection accuracy. 

\textbf{Why compare with full models:} Since all the previous approaches incorporate full neural network models, their detection processes are invariably completed either simultaneously with or after the inference of the full model, resulting in greater latency and energy consumption than the full model alone. Therefore, comparing our detection model with full models can effectively represents a comparison with previous methods.

\textbf{Results:} Table~\ref{tab:table8} illustrates that compared with the full model, the proposed lightweight detection model can significantly reduce latency across all the three platforms, achieving speedup range from 2.46$\times$ to 30.04$\times$. 
CPU$^{2}$ achieves the highest speedup with MobileNet-V2 at 30.04$\times$. 
Energy savings are significant, ranging from 55.27\% to 97.05\%. CPU$^{1}$ demonstrates consistent energy efficiency, while MobileNet-V2 on CPU$^{2}$ reaches the peak saving at 97.05\%. These findings show that the proposed method significantly enhances both latency and energy efficiency across various platforms compared to previous approaches. The performance on CPU$^{2}$ demonstrates that our method is particularly promising to be applied on edge devices.

Compared with previous approaches, our attention-based detection significantly reduces computational load by avoiding introducing costly operations like L-2 norm used in Yu~\cite{yu2021defending}. In terms of memory usage, our method requires only a single register to store one threshold during deployment, whereas LanCe's requires 1000 registers for 1000 thresholds~\cite{xu2020lance}. This difference results in a substantial reduction in memory utilization for our method.
Moreover, our lightweight detection model significantly reduces the complexity of implementation and hardware design, making it particularly suitable for resource-constrained environments.

%% file: _tab/tab1_v2.tex
\begin{table}[h]
    \vspace{-2mm}
    \setlength{\abovecaptionskip}{1.4mm}
    \setlength{\belowcaptionskip}{-0.cm}
    \small
    \caption{Top-1 Acc (\%) of DNNs Under Adversarial Attacks}
    \centering
        \tabcolsep 3.5pt
    \begin{tabular}{l|ccccc}
    \hline
        \tabincell{c}{Threat\\model}  &  \textbf{VGG-16} & \tabincell{c}{ResNet\\-50} & \tabincell{c}{MobileNet\\-V2} & \tabincell{c}{Inception\\-V3} & \tabincell{c}{ResNet\\-101} \\
    \hline
        Clean & 70.33 & 80.14 & 71.18 & 67.24 & 81.29 \\ \hline
        \tabincell{l}{AdvP}  &  \textbf{0.83} & 14.70 & 33.98 & 9.55 & 7.42 \\
    \hline
    \end{tabular}
    \vspace{-3mm}
    \label{tab:table1}
\end{table}

%% file: _tab/tab2.tex
\begin{table}[h]
    \vspace{-2mm}
    \setlength{\abovecaptionskip}{1.4mm}
    \setlength{\belowcaptionskip}{-0.cm}
    \small
    \caption{Detection Rate(\%) Comparison Using Different Shallow Layers of VGG-16 as Detection Layer}
    \centering
        \tabcolsep 5.5pt
    \begin{tabular}{l|cccc}
    \hline
        Layer & \textbf{Conv1} & \textbf{Conv2} & Conv3 & Conv4 \\ \hline
        Threshold & 1.55 & 4.48 & 2.91 & 7.59 \\ \hline
        AdvP-VGG16  & 97.28 & \textbf{97.36} & 97.34 & 97.42 \\ \hline
        AdvP-ResNet50 & 97.52 & \textbf{97.61} & 97.53 & 97.47 \\ \hline
        AdvP-MobileNetV2 & 88.64 & \textbf{97.53} & 89.01 & 55.47 \\ \hline
        AdvP-InceptionV3 & 97.44 & \textbf{97.54} & 97.43 & 97.33 \\ \hline
        AdvP-ResNet101 & 97.34 & \textbf{97.61} & 97.43 & 97.37 \\
    \hline
    \end{tabular}
    \begin{tablenotes}
        \footnotesize
        \item[] \textbf{Conv\#n}: Represents the cascade of shallow layers from the first layer to the n-th Conv layer of VGG-16 used as the detection model, which generates the necessary inputs for Equations (1).
        \item[] \textbf{AdvP-xxxx}: Represents an adversarial patch specially trained to attack the xxxx neural network.
        \item[] \textbf{Threshold}: Calculated from Equation (3) during the analysis stage and used to guide the deployment of the final comparison(as shown in Equation (4)) in real-world deployment.
     \end{tablenotes}
    \vspace{-6mm}
    \label{tab:table2}
\end{table}

%% file: _tab/tab3.tex
\begin{table*}[b]
    \vspace{-3mm}
    \setlength{\abovecaptionskip}{1.4mm}
    \setlength{\belowcaptionskip}{-0.cm}
    \small
    \begin{center}
    \begin{threeparttable}
        \centering
    \caption{Patches Detection Results with Shallow Layers of Other Models}

        \tabcolsep 9pt
    \begin{tabular}{l|cc|cc|cc|cc}
    \hline
        Model & \multicolumn{2}{c|}{ResNet-50} & \multicolumn{2}{c|}{MobileNet-V2} & \multicolumn{2}{c|}{Inception-V3} & \multicolumn{2}{c}{ResNet-101} \\
        Layer & \textbf{Conv1} & Conv2-1 & \textbf{Conv1} & Conv2 & Conv1 & Conv2 & Conv1 & Conv2-1 \\ \hline
        Threshold & 11.08 & 1.41 & 2.81 & 7.41 & 1.11 & 1.03 & 10.08 & 1.22 \\ \hline
         \tabincell{l}{AdvP\\-VGG16} & 97.65 & 97.51 & 97.74 & 80.24 & 97.44 & 97.57 & 97.55 & 97.30 \\ \hline
         \tabincell{l}{AdvP\\-ResNet50} & 97.50 & 97.58 & 93.90 & 80.06 & 97.21 & 97.70 & 97.47 & 97.34 \\ \hline
        \tabincell{l}{AdvP\\-MobileNetV2} & 84.71 & 50.19 & 83.20 & 67.56 & 57.42 & 57.76 & 64.02 & 52.09 \\ \hline
        \tabincell{l}{AdvP\\-InceptionV3} & 97.60 & 97.64 & 97.72 & 88.79 & 97.27 & 94.17 & 97.36 & 97.21 \\ \hline
        \tabincell{l}{AdvP\\-ResNet101} & 97.40 & 97.68 & 97.69 & 70.51 & 92.35 & 92.06 & 97.59 & 97.17 \\
    \hline
    \end{tabular}
    \begin{tablenotes}
        \footnotesize
        \item[] \textbf{Conv\#n}: Represents the cascade of shallow layers from the first layer to the n-th Conv layer of a specific DNN model.
     \end{tablenotes}
    \vspace{-3mm}
    \label{tab:table3}
    \end{threeparttable}
    \end{center}
\end{table*}

%% file: _tab/table8.tex
\begin{table*}[t]
    \vspace{-3mm}
    \setlength{\abovecaptionskip}{1.4mm}
    \setlength{\belowcaptionskip}{-0.cm}
    \small
    \begin{center}
    \begin{threeparttable}
        \centering
    \caption{Comparison of Latency and Energy.$\dagger$}     
        \tabcolsep 5pt
    \begin{tabular}{l|c|ccc|ccc}
    \hline
        \multirow{2}{*}{Model} & \multirow{2}{*}{Platform} & \multicolumn{3}{c|}{Latency (ms)} & \multicolumn{3}{c}{Energy (mJ)}\\ \cline{3-8}
        ~& ~ & \tabincell{c}{Detection (ours)} & Full model & Speedup & \tabincell{c}{Detection (ours)} & Full model & Energy saving\\ \hline
        VGG-16 & \multirow{3}{*}{GPU} & (Conv2)1.08 & 2.89 & 2.69 & (Conv2) 281.68 & 975.88 & 71.14\% \\
        MobileNet-V2 & ~ & (Conv1) 1.05 & 2.59 & 2.46 & (Conv1) 129.79 & 357.50 & 63.69\%\\
        ResNet-50 & ~ & (Conv1) 1.08 & 3.07 & 2.85 & (Conv1) 152.83 & 711.23 & 78.51\% \\ \hline \hline 
        VGG-16 & \multirow{3}{*}{CPU$^{1}$} & (Conv2) 9.48 & 58.56 & 6.18 & (Conv2) 994.7 & 7087.80 & 85.97\% \\
        MobileNet-V2 & ~ & (Conv1) 1.11 & 11.52 & 10.36 & (Conv1) 129.40 & 1029.36 & 87.43\% \\
        ResNet-50 & ~ & (Conv1) 2.05 & 33.00 & 16.07 & (Conv1) 192.94 & 2287.58 & 91.57\% \\ \hline \hline 
        VGG-16 & \multirow{3}{*}{CPU$^{2}$} & (Conv2) 25.83 & 98.79 & 3.82 & (Conv2) 175.41 & 392.18 & 55.27\% \\
        MobileNet-V2 & ~ & (Conv1) 5.12 & 153.87 & 30.04 & (Conv1) 33.15 & 1121.74 & 97.05\% \\
        ResNet-50 & ~ & (Conv1) 6.28 & 33.73 & 5.37 & (Conv1) 38.04 & 201.39 & 81.11\% \\ \hline
    \end{tabular}
    \begin{tablenotes}
        \footnotesize
        \item[] $\dagger$ The latency and energy are averaged per individual detection and inference sample.
        \item[] GPU: NVIDIA GeForce RTX 3090.   \quad
        CPU$^{1}$: 12$^{th}$ Gen Intel(R) Core(TM) i9-12900KF. \quad
        CPU$^{2}$: Apple M2
        \item[] Full model: the complete inference model of VGG-16/MobileNet-V2/ResNet-50.
     \end{tablenotes}
    \vspace{-8mm}
    \label{tab:table8}
    \end{threeparttable}
    \end{center}
\end{table*}

%% file: _txt/5-Conclusion_v1.tex
\section{Conclusion}
In this paper, we proposed a universal and efficient edge computing framework for robust AI. The framework leverages the attention map derived from the shallow layers' feature map of a DNN model to effectively identify a broad spectrum of visual attacks. The experiments conducted in this study have demonstrated the effectiveness and universal applicability of our approach, achieving an F-score of at least 97.4\% in detecting various attacks. Furthermore, our method offers several advantages for implementation on edge devices. It requires less computational cost and memory for registers and eschews introducing complex multiplication. The latency and energy consumption have been improved, achieving a speedup ranging from 2.46x to 30.04x and energy savings between 55.27\% and 97.05\%. These characteristics make it highly suitable for resource-constrained environments. Additionally, our approach does not require any knowledge of or modifications to the back-end AI model, enabling seamless integration with widely-deployed industrial AI models. 
Overall, our framework demonstrates the universality and efficiency of edge solutions for robust AI. 

